\documentclass[epj]{svjour}
\usepackage{graphics}

\def\snn{$\sqrt s_{\rm NN}$}

\newcommand{\beq}{\begin{equation}}
\newcommand{\eeq}{\end{equation}}    
\newcommand{\beqar}{\begin{eqnarray}}
\newcommand{\eeqar}{\end{eqnarray}} 
\newcommand{\ds}{\displaystyle} 

\begin{document}
\title{Directed flow in heavy-ion collisions at NICA: what is 
interesting to measure?}
\author{L.V. Bravina \inst{1,3,4} 
\thanks{e-mail: larissa.bravina@fys.uio.no}
\and E.E. Zabrodin \inst{1,2,3,4}}
\institute{Department of Physics, University of Oslo, Oslo, Norway \and
Skobeltsyn Institute of Nuclear Physics, Lomonosov Moscow State
University, Moscow, Russia \and
National Research Nuclear University "MEPhI" (Moscow Engineering Physics
Institute), Moscow, Russia \and 
Frankfurt Institute for Advanced Studies (FIAS), Goethe University,
Frankfurt a.M., Germany }
\date{Received: date / Revised version: date}
%
\abstract{
We study the formation of the directed flow of hadrons in nuclear 
collisions at energies between AGS and SPS in Monte Carlo cascade model. 
The slope of the proton flow at midrapidity tends to zero (softening) 
with increasing impact parameter of the collision. For very peripheral 
topologies this slope becomes negative (antiflow).
The effect is caused by rescattering of hadrons in remnants of the 
colliding nuclei. Since the softening of the proton flow can 
be misinterpreted as indication of the presence of quark-gluon plasma, 
we propose several measurements at NICA facility which can help one 
to distinguish between the cases with and without the plasma formation.
\PACS{
      {25.75.-q, 25.75.Ld, 24.10.Lx}{} 
     } 
} 
\maketitle
\section{Introduction}
\label{intro}

The collective flow of hadrons, produced in the course of nuclear
collisions at (ultra)relativistic energies, was found already quite 
long ago to be able to carry information about the energy density, 
pressure, compressibility, etc of the medium \cite{SMG74,SMG80,B83,SG86}.
Therefore, the flow was suggested as a good messenger of a new state 
state of matter - the quark-gluon plasma (QGP). At that time the 
anisotropic flow, appeared in non-central collisions, was subdivided
into two components \cite{SG86,ReRi97}, namely the {\it in-plane flow},
elongated within the reaction plane and called {\it bounce-off} or
{\it directed} flow, and the {\it out-of-plane flow}, orthogonal to 
the reaction plane and known as {\it side-splash} or {\it squeeze-out}
flow. The side-splash of the participants and the bounce-off of the
spectators were first detected by the Plastic Ball collaboration 
\cite{Gu84} at Bevalac (Berkeley). The measured average in-plane 
transverse momentum per nucleon, $\langle p_x(y) / A \rangle$, is found
to have a characteristic S-shape as a function of centrality. Its
magnitude increases with rising projectile/target mass and reaches a
maximum in semi-central collisions \cite{Gu84}. Both bounce-off and 
squeeze-out flows become stronger in the case of heavy fragments
\cite{Cser83}, which have relatively small undirected thermal 
velocities.

Note that at energies $E_{\rm lab}$ below few GeV the fraction of 
nucleons heavily dominates the spectrum of hadrons. It was found that 
the proton flow had a linear slope at midrapidity, and the scaling 
behaviour was observed, see \cite{Cser94} and references therein. - If 
one normalises the proton momentum $p_x$ and rapidity $y$ to the 
projectile center-of-mass momentum $p_{\rm proj}$ and rapidity 
$y_{\rm proj}$, respectively, and then determines the slope of the 
proton flow at midrapidity
\beq
\ds
F = \left. \frac{\partial \tilde{p_x}(\tilde{y})}{\partial \tilde{y}}
\right|_{\tilde{y}=0}~,
\label{eq1}
\eeq
where $\tilde{p_x} = p_x/p_{\rm proj}$ and $\tilde{y} = y/y_{proj}$, 
the reduced slopes $\tilde{F} = F/p_{\rm proj}$ appear to sit on the top 
of each other, thus revealing the scaling \cite{Cser94}. Violation of 
this scaling at collision energies of AGS and higher was predicted by 
one of the authors in \cite{Brav95}. Its origin will be discussed in 
detail in Sect.~\ref{sec3}.      
 
The new era in the investigation of collective anisotropic flow of
hadrons began about 20 years ago, when the Fourier decomposition of 
the azimuthal distribution of hadrons was proposed \cite{VoZh96}
\beq
\ds
\frac{d N}{d \phi} \propto 1 + 2 \sum \limits_{n=1}^{\infty}
v_n \cos{\left[ n (\phi - \Psi_n) \right] }~.
\label{eq2}
\eeq
Here $\phi$ and $\Psi_n$ denote the azimuthal angle between the particle 
transverse momentum and the participant event plane, and the azimuth of 
the event plane of $n$th flow component, respectively. The Fourier 
coefficients $v_n$ are the flow harmonics,
\beq
\ds
v_n = \langle \cos{\left[ n (\phi - \Psi_n)\right] }\rangle~,
\label{eq3}
\eeq   
which can be found after the averaging over all particles and all events.
Nowadays the first coefficients are colloquially known as {\it directed},
$v_1$, {\it elliptic}, $v_2$, {\it triangular}, $v_3$, flow, and so 
forth. It appears that the directed flow is still elongated in the 
reaction plane. However, in contrast to the former directed/bounce-off 
flow, $v_1$ is not just averaged projection of particle momentum on the 
impact parameter axis but rather its averaged ratio to the transverse 
momentum, $v_1 = \langle p_x/p_T \rangle$. Nonetheless it is clear
that, apart from the magnitude of the signal, the basic features of
the behaviour of former bounce-off and modern directed flow must be
similar.       

In case of the first order phase transition between the QGP and the 
hadronic matter, the disappearance of the pressure gradients in the
mixed plasma-hadrons phase sho\-uld lead to the softest point of the 
equation of state \cite{HuSh95}. Such a softening will be clearly seen 
in the excitation function of the directed flow \cite{Brach00,CsRo99}.
The STAR Collaboration has found  during the beam energy scan (BES) at 
RHIC that the slope of the proton flow at midrapidity changes the sign 
from positive to negative between 7.7 and 11.5~GeV \cite{v1_bes_star}.
After reaching a local minimum between 11.5 and 19.6~GeV, the slope 
tends to zero, though remains negative, with rising energy of the 
collision. Several models have tried to describe these observations by
invoking the hypothesis of a crossover type of quark-hadron phase 
transition \cite{KCIT14,IS15}.

In the present paper we argue that such a signal can be obtained by 
means of Monte Carlo cascade model, at least at a qualitative level. In 
Sect.~\ref{sec3}, therefore, we present several features and tendencies 
in the development of hadronic directed flow which are characteristic 
for the microscopic transport models. Various measurable signals are 
proposed to discriminate between the QGP state and pure hadronic states. 
We start from the short description of basic principles of the 
quark-gluon string model (QGSM) which is used for the calculations.  

\section{Quark-gluon string model}
\label{sec2}

The quark-gluon string model (QGSM) 
\cite{qgsm1a,qgsm1b,qgsm2a,qgsm2b} is a Monte Carlo
cascade model which utilises the Gribov Reggeon field theory (RFT)
\cite{RFT1,RFT2} and string phenomenology. It relies on the so-called
{\it topological} $1/N_c$ expansion \cite{Ven74} in quantum 
chromodynamics (QCD), where $N_c$ is the number of colours. For 
semi-hard hadronic collisions the perturbative QCD methods are employed 
\cite{Amprl91}. The model is similar to the dual parton model (DPM)
\cite{dpm} and the VENUS model \cite{venus}. It neglects the potential
interactions between hadrons and concentrates on hadronic rescattering.
QGSM takes into account (i) the processes with quark exchange and 
with quark annihilation, corresponding to Reggeon exchanges in GRT,
(ii) the processes with colour exchange, corresponding to the Pomeron
exchanges, (iii) single and double diffraction processes, (iv) elastic
scattering, (v) baryon-antibaryon annihilation, and (vi) hard 
gluon-gluon scattering with large momentum transfer. The model considers
secondary interactions of the produced hadrons with the spectators and
with the produced hadrons, as well as between some unformed ones 
(string-string interaction), based on the hadron formation time.
It provides a fair description of the global characteristics of $hh$,
$h$A and A+A collisions at AGS and SPS energies.

Although QGSM has no explicit assumption on the QGP formation, a lot of 
intermediate hadronic objects - strings - are formed during the initial 
stage of the collision. The strings absorb an essential part of energy
and baryon charge, thus these energetic non-hadronic objects could be
considered as non-equilibrated precursors of the QGP phase. Further 
details of the model can be found in 
\cite{qgsm1a,qgsm1b,qgsm2a,qgsm2b}. Directed flow
of hadrons in heavy-ion collisions at energies from AGS ($E_{\rm lab} = 
11.6$~AGeV) and up to RHIC (\snn=200~GeV) was studied within the QGSM in 
Refs.~\cite{Brav95,Amprl91,Bra94a,Bra94b,BZFF99,BFFZ00,ZFBF01,BCFFZ02,
ZBFF04,Bur05}. 
Many of the features in the development of the directed flow are not 
attributed solely to QGSM but are common for a wide class of MC cascade 
models. Let us discuss them. 
  
\section{Directed flow in QGSM}
\label{sec3}

\subsection{Softening of nucleon directed flow at midrapidity}
\label{subsec3_1}

\begin{figure}
\resizebox{0.5\textwidth}{!}{\includegraphics{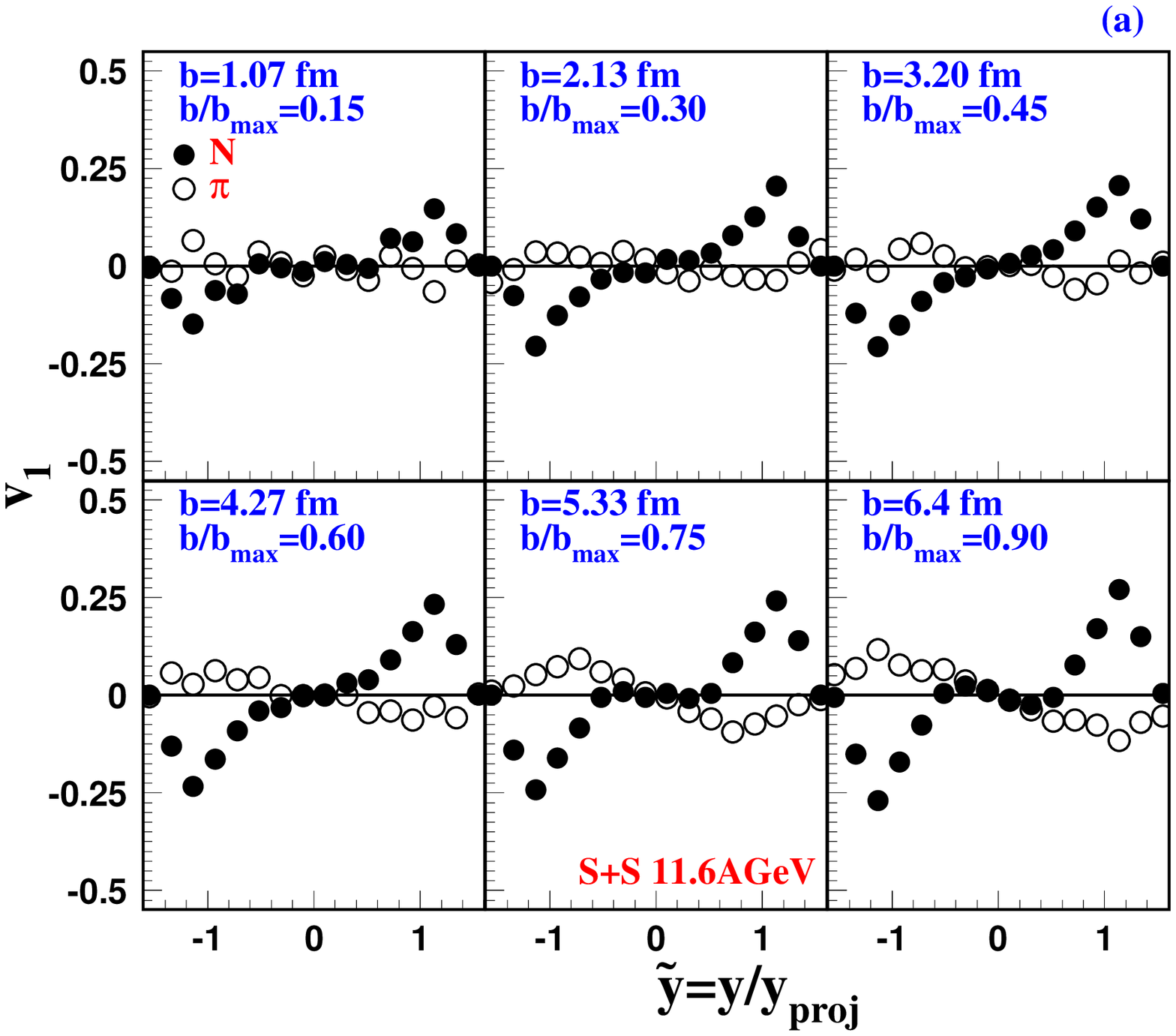}}
\resizebox{0.5\textwidth}{!}{\includegraphics{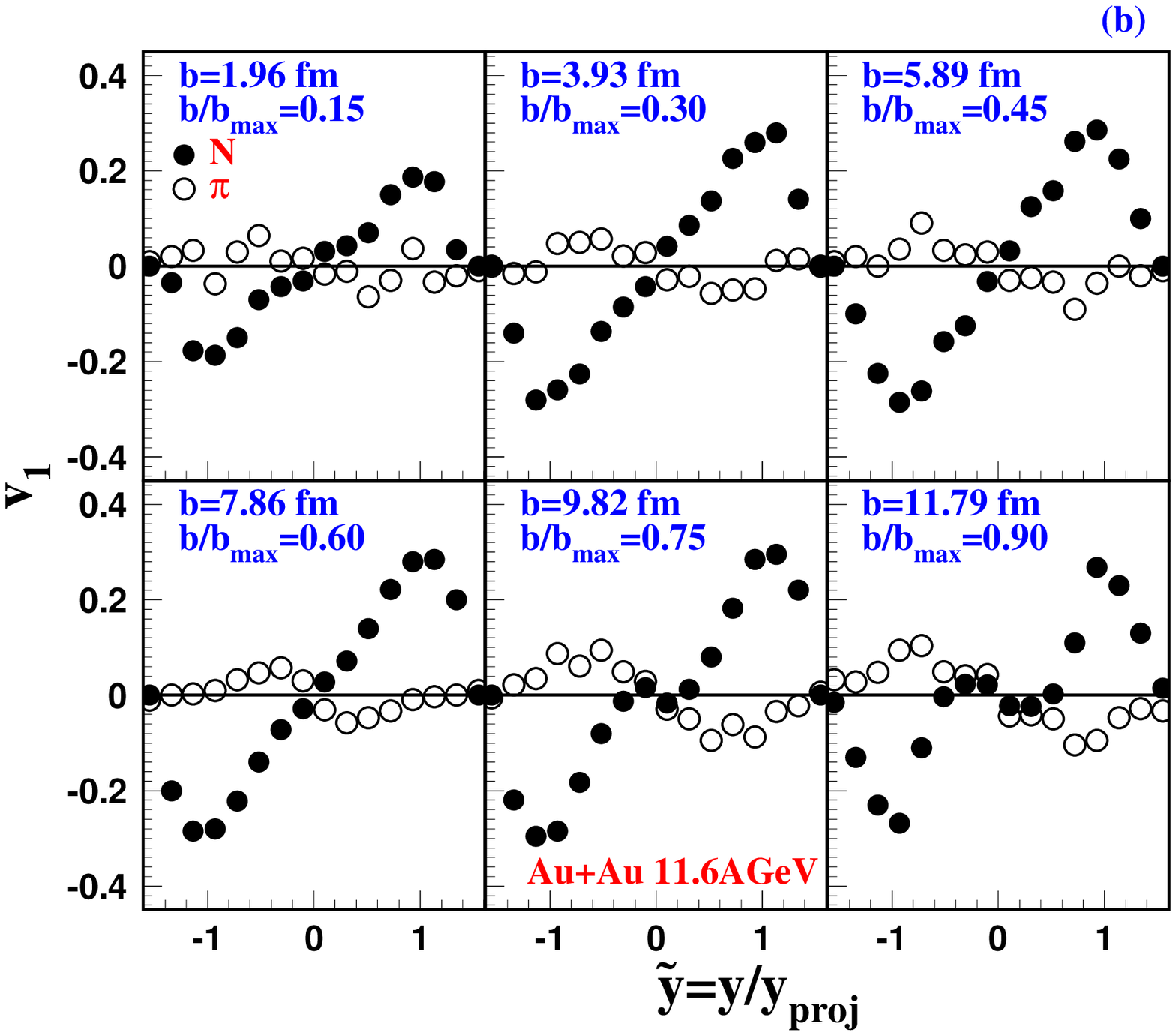}}
\caption{(Color online) 
(a) Directed flow of nucleons (full circles) and pions
(open circles) as a function of rapidity in $^{32}$S+$^{32}$S
collisions at 11.6{\it A} GeV/$c$.
(b) The same as (a) but for $^{197}$Au+$^{197}$Au collisions.
}
\label{fig1}       
\end{figure}

For better understanding of this phenomenon we choose collisions of 
light ($^{32}$S + $^{32}$S) and heavy ($^{197}$Au + $^{197}$Au) nuclei.
The comparison between the systems is done in terms of the reduced
impact parameter $\tilde{b} = b/b_{\rm max}$, where maximal impact 
parameter is equal to nuclear diameter, $b_{\rm max} = 2 R_A$, and 
reduced rapidity $\tilde{y} = y/y_{\rm proj}$. Results for 
$v_1(\tilde{y})$
distributions of nucleons and pions are shown in Fig.~\ref{fig1}(a,b)
for six centrality bins from $\tilde{b} = 0.15$ to $\tilde{b} = 0.90$.
For light system the slope of $v_1^N(\tilde{y})$ at midrapidity becomes
not so steep already at $\tilde{b} = 0.3$. The nucleon directed flow
decreases with rising $\tilde{b}$, and in peripheral collisions at
$\tilde{b} \geq 0.7$ it develops the antiflow similar to that of the 
pions. - Conventionally, we call the direction of the directed flow
{\it normal} if the product $p_x y$ is positive, $p_x y > 0$, and
{\it antiflow} if $p_x y < 0$. - The pion directed flow always 
demonstrates the antiflow behaviour. For heavy system the flatness of 
$v_1^N(\tilde{y})$ at midrapidity accompanied by the transition to 
antiflow seems to build up at $\tilde{b} \geq 0.7$ only. With increasing
energy of the collision,  however, this limit is shifted towards more 
central topologies \cite{BZFF99,ZFBF01}.
 
The reduction of the directed flow can be interpreted as softening of 
the EOS because of the QGP-hadrons phase transition \cite{HuSh95,CsRo99}.
One may also think about the colour field of the strings, which can 
lead to softening of hadronic EOS and thus imitate the plasma formation.
This hypothesis fails to explain why the effect of the $v_1^N$
disappearance is stronger (i) in collisions of light ions and (ii) in
peripheral collisions compared to the semi-central ones. The most 
reliable explanation of this phenomenon is the shadowing (or screening)
of emitted hadrons by the spectators. 

\begin{figure}
\resizebox{0.5\textwidth}{!}{\includegraphics{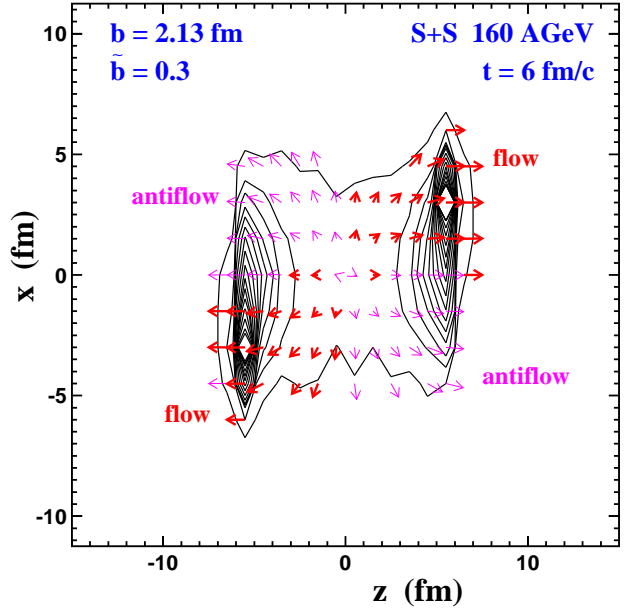}}
\caption{(Color online)
Space-time evolution of the baryon density (contour plots) and 
collective velocity (arrows) of the cells with volume $V = 3$ fm$^3$ 
each. Calculations are made for semi-central ($b = 2.13$ fm) S+S 
collisions at SPS energy for all formed baryons at time $t = 6$ fm/$c$. 
Contour plots correspond to $n_{\rm B} / n_0 = 0.05,\ 0.25,\ 0.5,\ etc$ 
of normal baryon density.
}
\label{fig2}
\end{figure}

To illustrate this idea we display in Fig.~\ref{fig2} a snapshot of 
collective velocities and baryon densities of the rectangular cells 
with volumes $V \simeq 3$~fm$^3$ in S+S collisions at SPS energy. The 
impact parameter is fixed at $b = 2.13$~fm corresponding to $\tilde{b} = 
0.3$, and the snapshot is taken at $t = 6$~fm/$c$ after beginning of the 
collision. The cells which develop normal flow always follow the 
baryon-rich spectator zones, whereas the antiflow is developed in more 
dilute areas. Hadrons with small rapidities, emitted early in the 
direction of normal flow, will be absorbed by flying spectators. These 
particles should get 
higher rapidities after a sequence of rescatterings. If the collision 
energy increases to that of RHIC or LHC, more hadrons are produced and 
the remnants of the nuclei leave the interaction zone much quicker, thus 
giving space for the development of flow in both directions. Directed 
flow at midrapidity will drop almost to zero, in agreement with the 
experimental observations \cite{v1_alice}. If the collision energy is 
the same, but instead of the light system collision of heavy ions takes 
place, then the number of produced hadrons increases. Here absorption 
of the early emitted hadrons by the spectators takes place as well,  
but this process becomes less efficient compared to that of the light 
system, where the isotropic particle radiation from the central part is 
not so strong. Therefore, the directed flow for heavy-ion systems 
displays softening in more peripheral collisions. At bombarding energies
below 1-2~GeV the directed flow of hadrons is dominated by the 
participant nucleons. The flow has a characteristic S-shape and is
stronger in heavier systems. Energies accessible for NICA lie in a very
interesting region, where transition from baryon-dominated matter to
meson-dominated one takes place. Recall, that at $E_{\rm lab} \approx 
40$~AGeV
fractions of baryons and mesons in heavy-ion collisions are about the
same. Somewhere here the antiflow of protons at midrapidity reaches its
maximum, whereas in the fragmentation regions the $v_1^p(y)$ turns back
to normal flow behaviour. Therefore, it is important to measure not 
only the midrapidity range $|y| \leq 1$, but also the areas of target 
and projectile fragmentation. Needless to say, that both central and 
peripheral collisions up to 80\% or more should be measured.     

\subsection{Directed flow of high-$p_T$ hadrons}
\label{subsec3_2}

This issue is also very interesting, and NICA's energy range suits well
for the study. First, let us present what was observed in the model 
calculations \cite{ZFBF01,BCFFZ02}. Here the directed flow of hadrons
was studied at SPS energy particularly in transverse momentum intervals
\beq
\ds
v_1(y, \Delta p_T) = \int \limits_{p_T^{(1)}}^{p_T^{(2)}}
\cos(n \Delta \phi) \frac{d^2 N}{dp_T^2 dy} dp_T^2 \left/
\int \limits_{p_T^{(1)}}^{p_T^{(2)}}
\frac{d^2 N}{dp_T^2 dy} dp_T^2 \right. \ .
\label{eq4}
\eeq 
It was found that the directed flow of nucleons in peripheral collisions
changed the slope at midrapidity from antiflow to normal flow provided 
the nucleons with transverse momenta $p_T \geq 0.6$~GeV/$c$ were 
selected. In addition, even directed flow of high-$p_T$ pions and kaons 
changed its sign and became positive. In microscopic model this
peculiarity is explained by early emission times of high-$p_T$ particles.
There is no sharp particle freeze-out in microscopic models, 
and hadrons are emitted continuously from the interaction zone albeit
with different production rates \cite{freeze_ags,freeze_sps}. At both 
AGS and SPS energies nucleons with highest transverse momenta are coming 
either from the initial nuclear splash at $t = 1$-2~fm/$c$ or later on 
from the expanding fireball. Pions in Au+Au collisions at AGS demonstrate 
the same tendency \cite{freeze_ags}, whereas the pions with highest $p_T$ 
in the reactions at energies of SPS or higher are produced only within 
the first 1-2~fm/$c$ \cite{freeze_sps} in primary inelastic 
nucleon-nucleon collisions. These pions carry normal directed flow.

In experiments, the sign flip of the directed flow of high-$p_T$ pions 
was observed in heavy-ion collisions at SIS (1~AGeV) \cite{Wag00},
AGS \cite{e877a,e877b}, and SPS \cite{na49} energies.
In line with the microscopic calculations, the effect was found to be
stronger in peripheral collisions and at higher rapidities. Since the
formation of rapidly expanding thermal source in light-ion collisions
or in peripheral heavy-ion collisions is less likely compared to more
central heavy-ion collisions, the most plausible explanation of the
effect is screening/shadowing by the remnants of the spectators. Note
also, that at SIS energies pions with high $p_T$ are emitted within
the first 15-20~fm/$c$ of the collision \cite{BHSW93,LBB91,Uma98}.
These pions experience at least 2-3 quasi-elastic scattering with 
nucleons, which lead to the partial thermalization of their spectrum
and, therefore, to possibility of their usage as a ``time-clock" for
probing the high-density phase. At SPS energies this option is quite
doubtful. Thus, the freeze-out of hadron species at NICA deserves 
further investigations.          

\section{Conclusions}
\label{concl}

The microscopic cascade model allows for reduction and changing the 
sign of the slope of directed flow of protons in (semi)peripheral 
nuclear collisions at relativistic energies. This effect can emulate 
the expected vanishing of the directed flow because of the phase
transition from QGP to hadrons, which causes the softening of the 
equation of state. In case of the QGP formation the softening of the
directed flow should be stronger in semi-central collisions and in
collisions of heavy ions compared to the light-ion ones. If the effect
is caused by hadron interactions in baryon-rich remnants of the 
colliding nuclei, the result should be opposite. To distinguish between 
the two phenomena, one can carry the following measurements at NICA 
facility:

\begin{itemize}
\item 
perform a run with light nuclei, say Cu+Cu or lighter. Results of the
flow measurements in heavy-ion and light-ion systems should be compared 
in terms of reduced impact parameter $\tilde{b} = b/2R_A$ and reduced 
rapidity $\tilde{y} = y/y_{\rm proj}$.

\item
measurement of directed flow in both central and peripheral events
(up to 90\% of centrality).

\item
measurement of particle $v_1$ not only at midrapidity but also in the
region of target/projectile fragmentation.

\item
determination of directed flow of hadrons with high, or even highest,
transverse momenta.  

\end{itemize}

\begin{acknowledgement}
L.B. acknowledges support of the Alexander-von-Humboldt Foundation.
\end{acknowledgement}


\end{document}